\pdfoutput=1
\documentclass[11pt,a4paper]{article}

\usepackage[T1]{fontenc}
\usepackage[utf8]{inputenc}
\usepackage{lmodern}
\usepackage{microtype}
\usepackage[margin=2.4cm]{geometry}

\usepackage{graphicx}
\usepackage{amsmath,amssymb}
\usepackage{mathtools}
\usepackage{upgreek}
\usepackage{array}
\usepackage{booktabs}
\usepackage{tabularx}
\usepackage{float}
\usepackage{enumitem}
\usepackage[font=small,labelfont=bf]{caption}
\usepackage{authblk}
\usepackage{xcolor}
\usepackage{url}
\usepackage[colorlinks=true,linkcolor=black,citecolor=blue!55!black,
            urlcolor=blue!55!black]{hyperref}

\newcolumntype{C}{>{\centering\arraybackslash}X}

\newcommand*\mystrutU[1]{\vrule width0pt height0pt depth#1\relax}
\newcommand*\mystrutO[1]{\vrule width0pt height#1 depth0pt\relax}

\title{\bfseries Oscillator-Based Processing Unit for\\ Formant Recognition%
\thanks{Accepted version of a paper published in \emph{Information} (MDPI), 2025.
Journal reference: \emph{Information} \textbf{2025}, \emph{XX}, XXXX;
DOI: \texttt{10.3390/XXXXXXXX}.}}

\author[1]{Tam\'as Rudner-Hal\'asz}
\author[2]{Wolfgang Porod}
\author[1,$\ast$]{Gy\"orgy Csaba}

\affil[1]{Faculty of Information Technology and Bionics, P\'azm\'any P\'eter Catholic University,
1083 Budapest, Hungary}
\affil[2]{Department of Electrical Engineering, University of Notre Dame (NDnano),
Notre Dame, IN 46556, USA}
\affil[$\ast$]{Corresponding author: \texttt{gcsaba@gmail.com}}

\date{}

\begin{document}
\maketitle

\begin{abstract}
\noindent
Oscillatory neural networks have been successfully applied to a number of computing
problems, such as associative memories and computationally hard optimization tasks.
In this paper, we show how to use oscillators to process time-dependent waveforms with
minimal or no preprocessing. Since preprocessing and first-layer processing are often
the most power-hungry steps in neural networks, our findings may open new doors to
simple and power-efficient edge-AI devices.

\vspace{0.8em}
\end{abstract}

%=====================================================================
\section{Introduction}

Oscillatory neural networks (ONNs) are analog circuits built from oscillatory building
blocks. They encode information in the phase, frequency and amplitude of these
oscillations and use oscillator--oscillator couplings to process data \cite{ref:perspectives,ref:aida}.
Given that oscillators are ubiquitous in the physical world and can be implemented with
very simple circuits, ONNs have the potential to become a practical circuit technology.
It has also been argued that oscillatory dynamics may naturally handle computationally
hard problems \cite{ref:herzog,ref:np2}.

There are a number of processing tasks where ONNs have been used successfully:
Hopfield-network-like associative memories \cite{ref:izhievich}, Ising machines
\cite{ref:herzog}, reservoir-like devices \cite{ref:velichko}, and they have even been
used as image classifiers with considerable success.

In many cases, ONNs do not need any training. This is the case, for example, in
oscillator-based Ising machines, where the problem statement itself defines the circuit
parameters. If training is needed, these networks mostly use simple Hebbian learning
rules to define their computing function \cite{ref:oschebbian}. Recently, we have shown
that it is possible to use gradient-based machine learning methods directly on the
circuit model of oscillators \cite{ref:frontiers}, which significantly extends the range
of problems that such oscillator networks can solve. It is also possible to use
equilibrium propagation \cite{ref:EP,ref:wang,ref:laydevant} or other novel algorithms
to create and train ONNs.

One of the main benefits of ONNs is that they offer the possibility of using very simple
building blocks: oscillators are ubiquitous in the physical world and can be realized by
just a few transistors (such as ring oscillators). However, the simplicity of ONN
processing is undermined by the need for input preprocessing. ONN-based associative
memories, for example, should receive an input oscillation with a stable phase, which is
not at all trivial to generate.

Preprocessing is also needed if ONNs are to be used on time-domain data. ONN-based
associative memories operate on images (i.e., on stationary patterns such as an image
encoded in a phase pattern), so time-domain data have to be converted to an image, such
as a spectrogram, first. In this way the rich oscillatory dynamics are likely not
exploited to their full extent.

In this paper, we attempt to alleviate the above-mentioned problems by using ONNs for the
classification of time-dependent signals with minimal preprocessing. Our case studies are
on speech (vowel) recognition.

Speech recognition is one example of a time-dependent processing problem. In most current
algorithms it is reduced to an image-processing task, as most conventional neural network
models solve it by using the mel spectrogram of the audio signal as the input to the
network \cite{ref:cetin}.

ONNs are dynamical systems with inherently rich dynamic behavior, so their capabilities
are not used to the full extent when only their stationary (converged) states enter the
computation. If a Hopfield-type classifier is realized by an ONN, for example, then a
stationary input (an input image or pattern) is presented and, at the end, a stationary
phase configuration is read out. The dynamics of the system matter only insofar as they
ensure convergence to a well-defined computational ground state.

When ONNs are used to process time-dependent signals, this is usually done within the
framework of reservoir computing \cite{reservoir1,reservoir2}. The main reason is likely
that dynamical systems are difficult to train, whereas reservoirs do not require training
of the said dynamical systems.

The novelty of our paper is that we follow a different route, attempting the design of
ONNs that do not rely on reservoirs. While reservoirs have a complex, unknown internal
structure, the ONN-based layers we design act as filters that extract or classify
spectral features of the input signals.

In this paper, we design two types of ONN architecture in which a first layer of
oscillators does the lion's share of the processing of the input auditory signals. This
first layer can directly take a time-domain signal and may output either a classification
result or a slowly time-varying signal that can be processed further by a traditional
neural network.

The first ONN architecture consists of a hidden layer with oscillator frequencies tuned to
the dominant frequency components of the incoming signals. The inputs of this ONN are the
formants of the vowels to be classified. Due to injection locking, groups of oscillators
may synchronize to each other (and to the input), signaling the presence of certain
components in the input. The synchronizing oscillators in the first layer may in turn
induce synchronization in the subsequent layer, indicating the joint presence of
characteristic frequency components. We show that the coupling strengths and the
frequencies of the oscillators can be optimized by machine learning methods. For certain
vowel classes, we reach high (close to 100\%) classification accuracy with the oscillators
alone.

The second ONN we demonstrate is a single-layer ONN that needs no preprocessing at all:
instead of receiving formant frequencies, it receives vowel waveforms directly, and groups
of oscillators respond to such stimuli. This ONN needs additional layer(s) of traditional
neural networks for a high-accuracy output, but we will demonstrate that the ``heavy
lifting'' is still done by the ONN layer.

Both architectures belong to the class of frequency-based ONNs, where the input and the
output of the computation are encoded in frequencies instead of phases
\cite{ref:suppes,ref:nikonov}. Such frequency-based schemes arise in various
oscillator-based computing fields, such as computational neuroscience \cite{ref:spike},
and in mathematical oscillator models such as modified Kuramoto oscillators
\cite{ref:taylor}.

It is important to emphasize that the goal of our work is not to advance the state of the
art in neural network accuracy, or to compete with large-scale models containing millions
of parameters. Our aim is rather to demonstrate that oscillator-based systems can serve as
simple, energy-efficient front-end layers in neural networks. Such systems have the
potential to reduce or even eliminate the need for components such as analog-to-digital
(A/D) converters and spectrum analyzers.

Our work paves the way for the direct processing of dynamic, time-domain signals. This is
likely to result in significant energy savings: in a typical edge-AI architecture the
first layer is the one that processes the raw data, and it is also responsible for the
bulk of the task at hand, so energy savings in this layer are the most impactful.

%=====================================================================
\section{Materials and Methods}

\subsection{Simulation of Coupled Ring Oscillators}

Ring oscillators are among the simplest oscillators. They are built from an odd number of
inverters connected in a ring \cite{ref:frontiers}. They are easy to implement, but they
possess inherent nonlinearities that make them suitable for complex, dynamic computing
tasks.

Ring oscillators can be coupled together through one of their nodes and, depending on
which nodes are coupled, the resulting pair is either positively or negatively coupled
\cite{ref:csabatrond}. The former means that the oscillators try to align in phase, the
latter the opposite, so that they align in an anti-phase state. In both cases their
frequencies pull towards each other. Since we are interested in a frequency-based ONN
(i.e., phases are not directly relevant), we only use positive couplings.

In our computational study we use a differential-equation-based (ODE-based) model of ring
oscillators, following the formalism of \cite{ref:frontiers}. This gives results very
closely matching a behavioral SPICE simulation model.

Several oscillators coupled together can be described by the following ODE, as presented
in \cite{ref:frontiers}:
\begin{equation}\label{eq:ode}
    C\frac{d\boldsymbol{v}}{dt} =
        \underbrace{\mystrutU{3.5ex}\frac{1}{R}\,\boldsymbol{A}'
            \big(f(\boldsymbol{P}_{\pi}\boldsymbol{v}) - \boldsymbol{v}\big)}_{\text{Intrinsic dynamics}}
        +
        \overbrace{\mystrutO{4.0ex}\frac{1}{R_{c}}\,\boldsymbol{C}'\boldsymbol{v}}^{\mathclap{\text{Coupling dynamics}}}
        -
        \underbrace{\mystrutU{3.5ex}\frac{1}{R_{\mathrm{in}}}
            \Big(\boldsymbol{B}'\odot(\boldsymbol{v} \ominus \boldsymbol{u}^{T})\Big)\boldsymbol{1}}_{\text{External dynamics}}
\end{equation}

\begin{figure}[htbp]
    \centering
    \includegraphics[width=\textwidth]{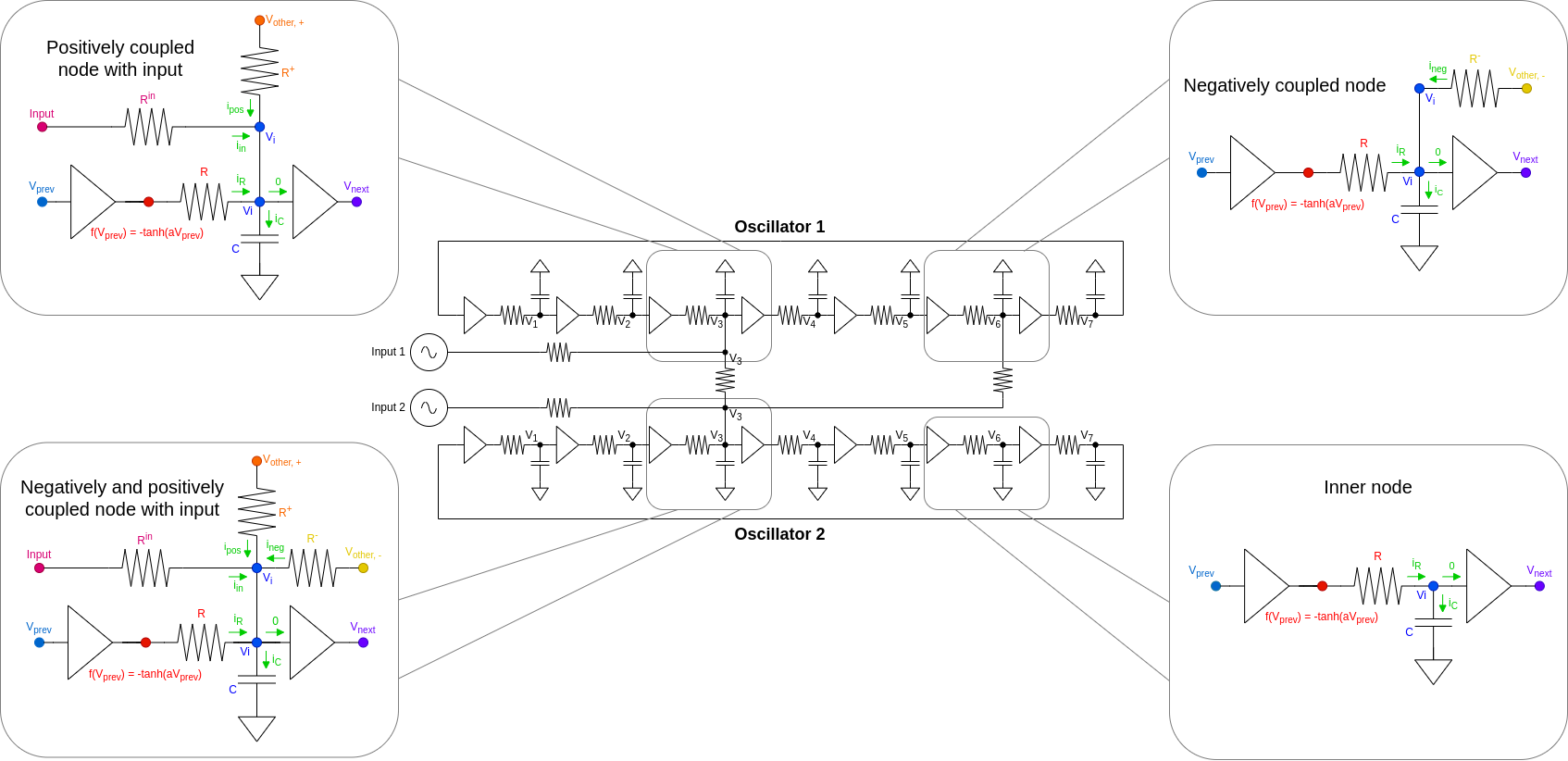}
    \caption{The architecture of a two-oscillator system. The oscillators are coupled both
    positively (node 3--3 connection) and negatively (node 6--3 connection). The different
    types of node, in terms of connectivity, that can occur in a ring oscillator are
    highlighted on the sides. For generality, the figure shows both positive (in-phase)
    and negative (anti-phase) couplings, but only positively coupled oscillators are used
    in this work.}
    \label{fig:ODE}
\end{figure}

This ODE is derived from Kirchhoff's laws; for the meaning of the individual terms, see
Figure~\ref{fig:ODE}. Here $\boldsymbol{P}_{\pi}$ is a permutation matrix that assigns the
node voltages according to the ring structure. $\boldsymbol{A}'$ is a diagonal matrix used
to set the frequency of the oscillators by modifying the resistance between the inverters
of a given oscillator. $\boldsymbol{C}'$ and $\boldsymbol{B}'$ are coupling matrices,
describing the oscillator--oscillator couplings and the couplings between the input
generators (denoted by $\boldsymbol{u}$) and the oscillators, respectively. The function
$f()$ describes the inverter nonlinearity. In this particular equation the input
generators are voltage generators, but current generators can also be used with a slightly
modified ODE.

Any ring-oscillator-based architecture can be designed by determining the
$\boldsymbol{A}'$, $\boldsymbol{B}'$ and $\boldsymbol{C}'$ matrices, which physically
amounts to setting the values of the resistances. This can be done either by a machine
learning method based on some learning criterion or, for smaller systems, by trial and
error ``by hand''.

\subsection{Frequency-Based Computing}

Our frequency-based computing scheme rests on the simple fact that two oscillators
synchronized to a common driving signal become synchronized to each other as well --- a
phenomenon well known in the mathematical literature on synchronization
\cite{pikovsky,ref:izhievich}. Certain frequency components of the driving signal can
therefore be detected by detecting the mutual synchronization of otherwise independent
oscillators.

\begin{figure}[htbp]
    \centering
    \includegraphics[width=0.95\textwidth]{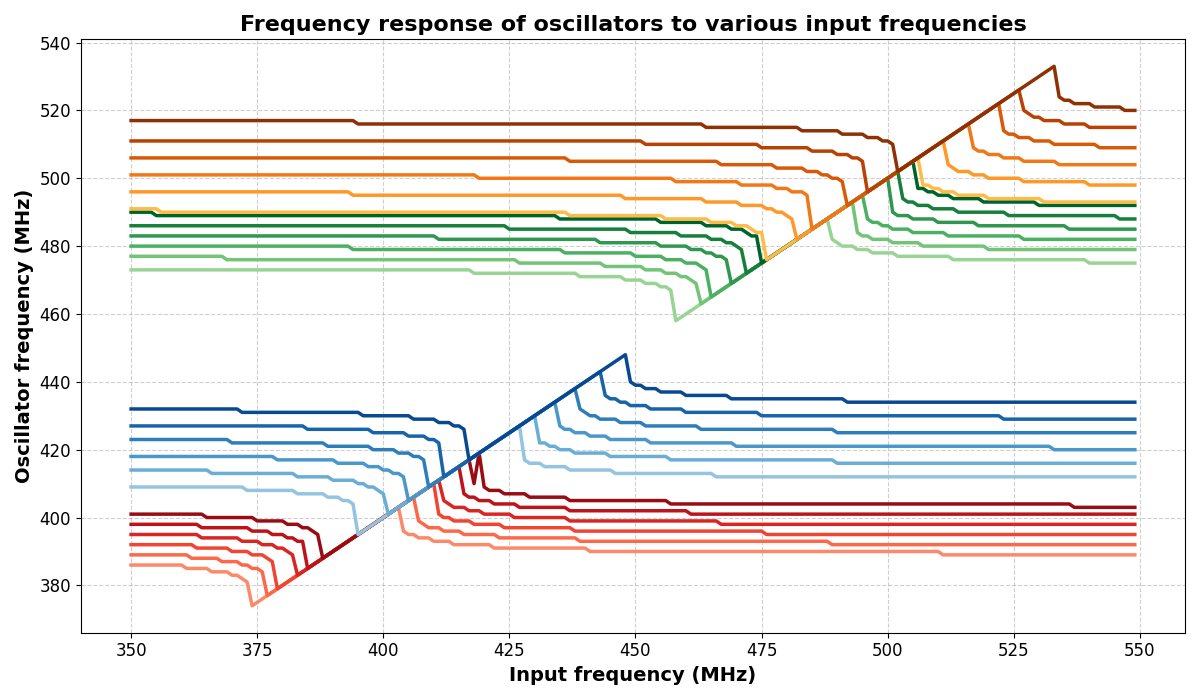}
    \caption{Frequency synchronization of oscillators to an input sinusoid. The frequency
    of a single input sinusoidal voltage generator, connected to a network of 24 uncoupled
    oscillators, is swept over a range of frequencies. The four groups of oscillators
    either synchronize to the input signal or not, depending on the frequency of the
    input. Within each group the oscillators synchronize to the input frequency one by
    one, until all of them are synchronized; they then run together over a range of
    frequencies, after which they depart from the group one by one again.}
    \label{fig:freq}
\end{figure}

If a sinusoidal signal arrives at one of the nodes of an oscillator, the oscillator is
capable of synchronizing its frequency to the frequency of the input signal, provided that
the latter is sufficiently close to the free-running frequency of the oscillator. This
interaction is illustrated for several oscillators in Figure~\ref{fig:freq}. The
oscillator frequencies are grouped, so the figure contains four groups of six oscillators,
with frequencies close to each other within a group but relatively far apart between
groups. The frequency of an input sinusoidal voltage generator is swept over the range
shown on the $x$ axis. When the input frequency gets close to the frequency of a group of
oscillators, all the oscillators in that group jump together (synchronize) in frequency.

\subsection{Vowel Database and Preprocessing}

To test our ONN system, we used a relatively simple, well-tested set of spoken English
vowels, as described in \cite{ref:hughes}.

The database comprises 12 vowels: ``ae'', ``ah'', ``aw'', ``eh'', ``ei'', ``er'', ``ih'',
``iy'', ``oa'', ``oo'', ``uh'' and ``uw''. The recordings are at most 1~s long and were
sampled at 16~kHz; we padded all of them with zeros to a length of exactly 1~s. The
dataset contains recordings by both men and women.

\begin{figure}[htbp]
    \centering
    \includegraphics[width=0.72\textwidth]{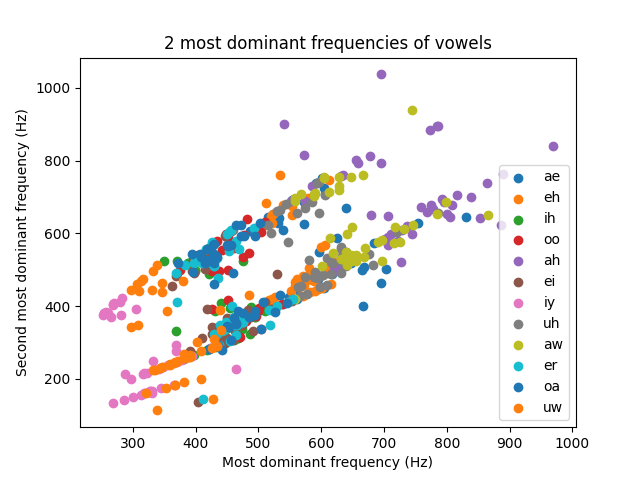}
    \caption{Scatter plot of the first and second most dominant frequency components of all
    the vowels in the dataset. The two lines along which the data points gather possibly
    arise from the fact that these vowels were recorded by both male and female speakers,
    and female speakers tend to have a higher pitch.}
    \label{fig:all}
\end{figure}

These vowels have two to four dominant frequency components, but in most cases the two
most dominant ones are pronounced enough to be extracted and used as preprocessed,
synthetic sinusoidal inputs to an oscillator-based architecture. A scatter plot of the
first two dominant frequency components of all the vowels is shown in
Figure~\ref{fig:all}. The two distinct bands along which the data points are scattered
--- the two roughly parallel lines --- most likely arise from gender differences between
the speakers.

For our simulations we used both the raw waveforms and the frequency components extracted
from them. In each case the time axis was scaled so that the vowel frequencies, originally
in the $100$--$1100$~Hz range, were shifted to the $100$--$1100$~MHz range. This scaling
was necessary in order to obtain realistic parameters for the oscillator network, bringing
its frequencies into the high-MHz to low-GHz range.

\subsection{Vowel Recognition by Frequency-Based Computing}

The key to the operation of our scheme is that oscillator groups detect frequency
components in the input waveform. This phenomenon can be used to detect audio signals
composed of a small number of dominant frequency components, such as English vowels.

A key requirement for the operation of this scheme is the proper choice of the oscillator
frequencies, which must match typical vowel frequencies. These frequencies are learnable
parameters that can be determined either by gradient-based or by gradient-free
optimization techniques.

\subsection{Optimization of Parameters}

Optimizing the ONNs means finding the right parameters in the $\boldsymbol{A}'$,
$\boldsymbol{B}'$ and $\boldsymbol{C}'$ matrices of Equation~\eqref{eq:ode}. In the
simplest case this can be done by trial and error --- the frequencies, for example, can be
estimated from Figure~\ref{fig:all}. Other methods can also be used, such as
backpropagation through time (BPTT) \cite{ref:torchdiffeq} or gradient-free optimization.
We used the trial-and-error and the gradient-free approaches, since we have shown earlier
that BPTT for ONNs is both memory-intensive and slow \cite{ref:frontiers}. Gradient-free
optimization, by contrast, frees us from the burden of a method with a heavy computational
cost.

\subsubsection{Gradient-Free Optimization}

Gradient-free optimization is often used when computing the derivative of the loss
function is costly or even impossible. The underlying idea is the same as for
gradient-based methods: the goal is to optimize some cost or loss function. There are many
variants of this kind of optimization, such as evolutionary algorithms. We used the
Nevergrad Python package \cite{ref:nevergrad} to find the free-running frequencies of the
oscillators.

\subsection{Effect of Time-Domain Vowel Signals on Oscillators}
\label{sec:voweleffect}

In the scenario of Figure~\ref{fig:freq}, the oscillators are subjected to a coherent,
sinusoidal injection signal. Waveforms to be classified are typically far from this ideal:
most time series are wideband signals with continuously and abruptly shifting frequencies,
and many different frequencies typically act on a particular oscillator simultaneously.

In the circuit model, an arbitrary waveform (such as a vowel waveform) can be applied to
the oscillators in a straightforward manner. We did this by connecting an additional
voltage generator through a resistor to a ring oscillator node.

\begin{figure}[htbp]
\centering
    \begin{minipage}[b]{0.8\textwidth}
        \centering
        \includegraphics[width=\linewidth]{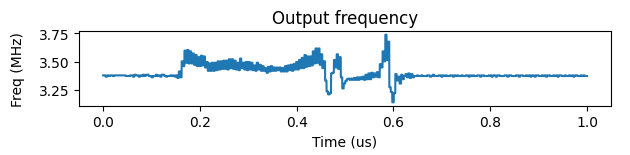}\\[2pt]
        (\textbf{a})
    \end{minipage}

    \vspace{0.4cm}

    \begin{minipage}[b]{0.8\textwidth}
        \centering
        \includegraphics[width=\linewidth]{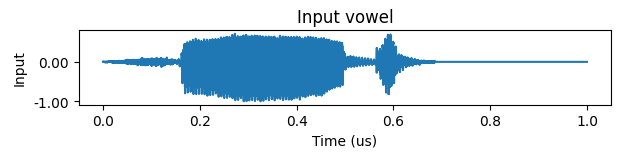}\\[2pt]
        (\textbf{b})
    \end{minipage}

    \vspace{0.4cm}

    \begin{minipage}[b]{0.8\textwidth}
        \centering
        \includegraphics[width=\linewidth]{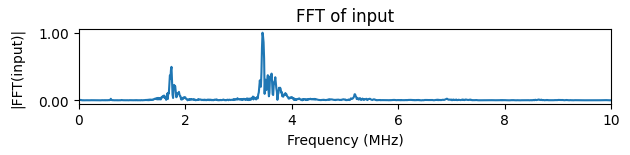}\\[2pt]
        (\textbf{c})
    \end{minipage}

    \caption{Effect of a time-domain waveform on a single oscillator.
    (\textbf{a}) Instantaneous frequency of an oscillator that is driven by the vowel
    waveform through a voltage generator.
    (\textbf{b}) The vowel in the time domain.
    (\textbf{c}) The Fourier spectrum of the vowel; the dominant frequency component of
    this particular vowel is around 350~MHz. Note that the vowel has been sped up, and
    hence shifted in frequency, by applying it over $1\,\upmu$s instead of 1~s. The
    frequency of the oscillator changes while the vowel is spoken on the recording and
    synchronizes to about 350~MHz, after which it returns to its free-running value of
    about 330~MHz. This shows that ring oscillators are capable of reacting to external
    stimuli with their frequency, even if the input is not coherent.}
    \label{fig:vowel}
\end{figure}

The effect of a vowel waveform on an oscillator is illustrated in
Figure~\ref{fig:vowel}. The waveform itself is given in Figure~\ref{fig:vowel}b and its
Fourier spectrum in Figure~\ref{fig:vowel}c. As described above, the sampling frequency
was scaled so that the spectrum falls close to the free-running frequency of the
oscillator.

Figure~\ref{fig:vowel}a shows the instantaneous frequency of the oscillator, determined
simply from the period of the oscillation. The frequency of the oscillator is continuously
pulled towards the dominant frequency of the waveform for as long as the vowel waveform
acts on the oscillator.

When the vowel is actually spoken in the recording, the frequency of the oscillator
changes and synchronizes to the principal frequency of the input vowel for the duration of
the vowel; when the speaker stops, it returns to the free-running frequency of the
oscillator.

The oscillator may therefore be viewed as a ``filter'': its frequency is pulled whenever a
nearby frequency component appears in the input. Importantly, this effect survives even
for a highly incoherent input.

\begin{figure}[htbp]
    \centering
    \includegraphics[width=0.8\textwidth]{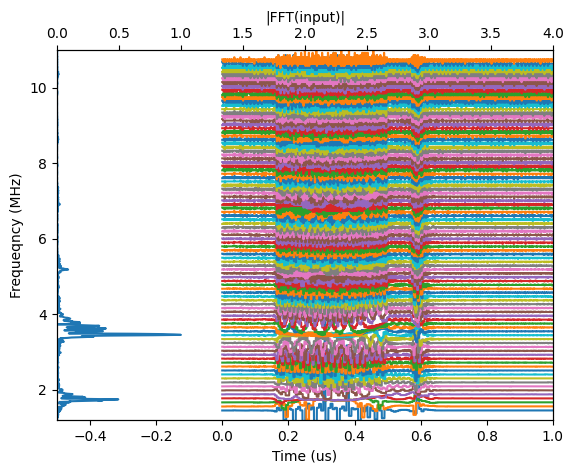}
    \caption{Frequency response of a layer of oscillators to a single, time-domain input
    vowel. Oscillators with free-running frequencies close to the peaks of the input
    spectrum synchronize in frequency to each other for as long as the vowel is spoken,
    and then return to their original frequencies. Colors are used only for better
    visibility.}
    \label{fig:layer}
\end{figure}

To use this effect for vowel recognition, we apply the vowel waveform simultaneously to a
number of oscillators whose frequencies span the spectral range of the vowel. The behavior
of such an oscillator filter bank for one particular input is shown in
Figure~\ref{fig:layer}, where again the instantaneous frequencies are plotted as a
function of time. Groups of oscillators with closely lying frequencies jump together when
the vowel waveform acts on them. Oscillators with frequencies further away are affected by
the vowel but do not synchronize in frequency. It is important to note that oscillators
that frequency-synchronize in this scenario are typically not phase-synchronized, and
their waveforms do not overlap.

Qualitatively, the oscillator layer may be viewed as a filter bank in which the presence
of a particular frequency component is signaled by the synchronization of the oscillators
around that frequency.

\subsection{Integrating the ONN Layer with Traditional Neural Networks}

For some of the results below we used the ONN in tandem with traditional layers, which
serve to improve the results coming from the oscillator layers.

For the training of the conventional multi-layer perceptron introduced in the next
section, we used standard methods: the Adam optimizer and a cross-entropy loss,
implemented in PyTorch \cite{ref:pytorch}. Because this added network is small, training
itself was fast and lightweight.

%=====================================================================
\section{Results}

In this section we introduce two architectures for vowel classification.

The first architecture is a two-layer neural network made entirely of oscillators. It uses
two of the extracted dominant frequency components (formants) of the spoken vowels as
inputs.

The second architecture uses an ONN as its first layer, with the classification performed
in a second, conventional (perceptron-type) layer. It takes the vowel waveforms directly
as inputs, with no preprocessing.

Both architectures are tested on a binary classification problem, in order to show that
they are capable of solving such tasks accurately; we picked the ``ah'' and ``iy'' vowels
as our test case.

\subsection{Vowel Recognition by a Multi-Layer Oscillatory Neural Network}
\label{sec:multilayer}

\begin{figure}[htbp]
    \centering
    \includegraphics[width=\textwidth]{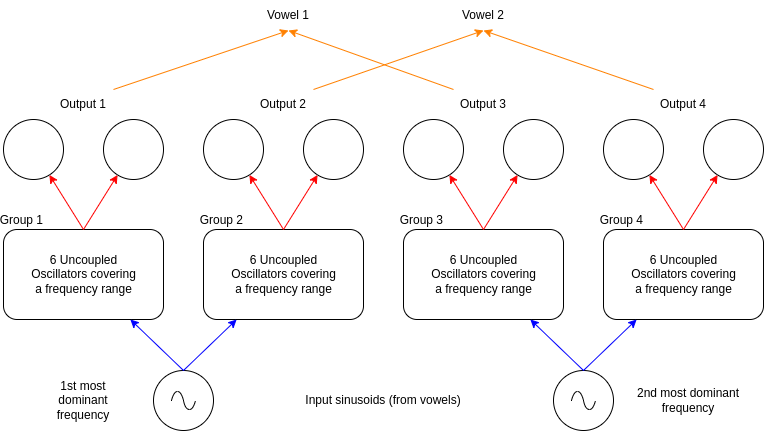}
    \caption{Our proposed architecture. The input signals are represented by sinusoidal
    voltage generators whose frequencies are set to the first and second most dominant
    frequency components of a given input vowel. These are connected to a single layer of
    uncoupled oscillators, grouped into four distinct groups to highlight their collective
    responsibility for detecting a certain frequency range in the input signal. A group
    synchronizes in frequency if an input sinusoid falls within the frequency range that
    the group covers. The groups are in turn connected to a pair of oscillators covering
    the same frequency range: if the six oscillators of a group are synchronized in
    frequency, the corresponding pair in the output layer will also be synchronized in
    frequency. In this way the presence of particular frequencies in the input sinusoids
    is signaled by different output pairs.}
    \label{fig:arch}
\end{figure}

The first architecture we propose, which distinguishes two vowels on the basis of their
two most dominant frequency components, is composed of two layers and is shown in
Figure~\ref{fig:arch}.

The two oscillator layers have no intralayer couplings, so no oscillator in a given layer
is connected to any other oscillator of the same layer.

The two layers serve two distinct purposes. The first layer detects the frequency
components of the input signal; the frequency synchronization within a group signals the
presence of a given frequency range in the input. The second layer performs the actual
classification step, by detecting the pair of frequency components that is characteristic
of the vowel.

First, the two frequency values are turned into two distinct sinusoidal signals with the
given frequencies. These signals are then fed to the corresponding groups of the first
layer: the first dominant frequency component drives Groups 1 and 2, and the second
drives Groups 3 and 4.

The oscillator groups in the first layer of Figure~\ref{fig:arch} are used to detect the
presence of frequency components. Since we use two frequency components in a binary
recognition task, there are four possible frequency combinations, and the task of the
first layer is to signal whether these four frequency ranges are present in the input
sinusoids or not.

This frequency detection is then fed forward to the second layer. Each of the four groups
of the first layer is connected to two oscillators of the output layer. The mechanism in
the second layer is simple: if the oscillators of a group in the first layer are
synchronized in frequency, then the two oscillators of the second layer to which they are
connected will also synchronize in frequency. In this way, a simple frequency measurement
in the second layer suffices to determine which frequency ranges were present in the input
sinusoids.

This is a very simple architecture with a low number of connections. The numbers of
couplings and oscillators are summarized in Table~\ref{tab:param_counts}.

\begin{table}[htbp]
    \centering
    \caption{Neuron, coupling and parameter counts for the proposed multi-layer
    oscillatory network. The network is lightweight, with only 50 couplings. Note that the
    number of quantities the network has to determine is the sum of the number of
    parameters and the number of couplings: the former set the frequencies of the
    oscillators, the latter the strengths of the couplings between connected oscillators.}
    \label{tab:param_counts}
    \begin{tabularx}{\textwidth}{lCCCC}
        \toprule
         & \textit{\textbf{Input}} & \textit{\textbf{Layer \#1}} & \textit{\textbf{Layer \#2}} & \textbf{All} \\
        \cmidrule{2-5}
        \textit{\# of neurons}                & 2 & 24 &  8 & 34 \\
        \textit{\# of parameters}             & 0 & 24 &  8 & 32 \\
        \textit{\# of couplings to next layer} & 2 & 48 &  0 & 50 \\
        \bottomrule
    \end{tabularx}
\end{table}

\begin{figure}[htbp]
    \centering
    \includegraphics[width=0.7\textwidth]{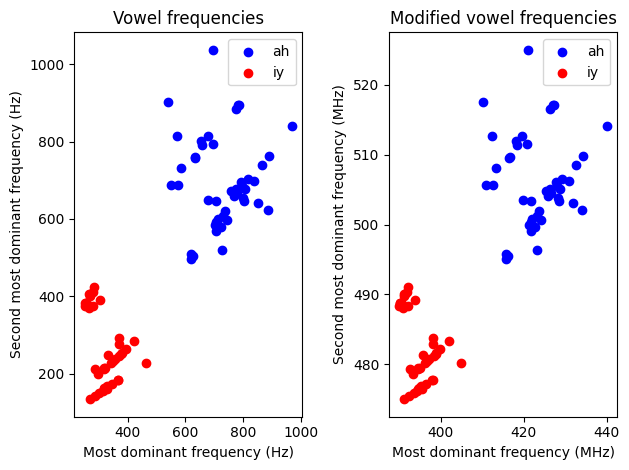}
    \caption{The dataset, consisting of the two vowels ``ah'' and ``iy'', that we used to
    test the proposed architecture. This is a linearly separable problem, so the
    architecture should have no difficulty solving it.}
    \label{fig:binary}
\end{figure}

To test the performance of the network, we extracted the dominant frequency components of
the two vowels mentioned above. A scatter plot of these data points is shown in
Figure~\ref{fig:binary}, which contains both the baseline frequencies and the scaled
frequencies.

It is worth noting that this network is expected to classify correctly only linearly
separable problems, which is the case for the vowels shown in Figure~\ref{fig:binary}.

\subsubsection{Choosing the Network Parameters for High-Accuracy Classification}
\label{sec:params}

In order to configure the network of Figure~\ref{fig:arch}, the frequencies of the
oscillators and the coupling strengths --- between the input voltage generators and the
hidden layer, and between the hidden and the output layers --- have to be set. The
coupling strength translates physically into a resistance value. In this section we show
the results of setting these parameters by simple considerations.

Since the problem involves four distinct frequency ranges --- vowels being differentiated
on the basis of their primary and secondary dominant frequency components --- configuring
the frequencies of the hidden and output layers is straightforward. We only needed to
determine the frequencies of the grouped oscillators that interact with the input
sinusoids, and to select a coupling strength between the oscillators that would neither
distort the signals nor cause the synchronization regions to overlap.

\begin{figure}[htbp]
    \centering
    \includegraphics[width=0.8\textwidth]{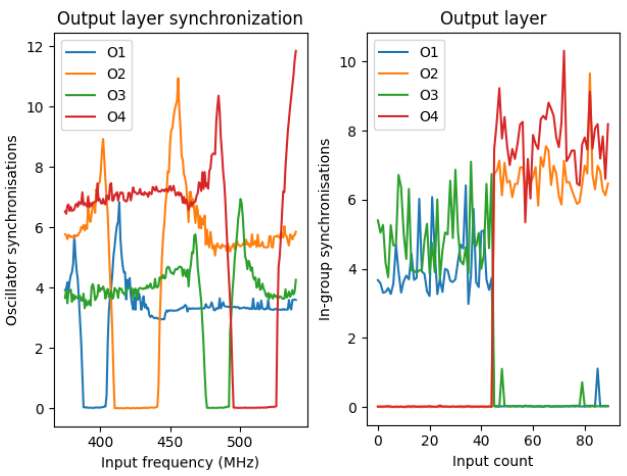}
    \caption{Left: the synchronization value of the four pairs of output oscillators. The
    frequency of the input sinusoid is swept over a given frequency range, and
    synchronization appears as the flat bottom of the well in each curve. Right: the same
    synchronization value, calculated after feeding the two types of vowel to the system.
    For the first 45 vowels, which are ``ah'', the O2 and O4 pairs are synchronized; for
    the other 45 vowels, which are ``iy'', the O1 and O3 pairs are synchronized.}
    \label{fig:ring}
\end{figure}

The left panel of Figure~\ref{fig:ring} shows the synchronization regions of the
two-oscillator groups of the output layer as wells in the individual curves. The $x$ axis
is a sweep of the input frequency over an extended range that covers the two most dominant
frequency components of the vowels. The O1 and O2 groups are responsible for recognizing
the lower frequencies and the O3 and O4 groups the higher ones, and the synchronization
regions do not overlap.

In order to measure and decide algorithmically whether a group of oscillators is
synchronized or not, we simply used the following measure:
\begin{equation*}
    S = \sum_{i, j} (f_i - f_j)^2,
\end{equation*}
where the $f_i$ are the frequencies of the oscillators in the given group. This $S$ value
is then scaled and used as a dimensionless number, because the magnitude of the difference
is not relevant --- only the fact that it is low for synchronizing and high for
non-synchronizing cases.

The right panel of Figure~\ref{fig:ring} shows the synchronization values of the four
oscillator groups of the output layer as a function of the input index. The first 45
samples are the ``ah'' vowel and the remaining 45 the ``iy'' vowel. It is clear that with
this synchronization measure the decision is easy: if the sum of the synchronization
values of O2 and O4 is lower than that of O1 and O3, the input is an ``ah'' vowel;
otherwise it is an ``iy'' vowel. For the first batch of samples the sum for O2 and O4 is
close to zero while the sum for the other two is high, and for the second batch, with the
other vowel, it is the other way around.

From this we can also conclude that the architecture is capable of solving this task with
$100\%$ accuracy.

The physical parameters of the network are summarized in
Table~\ref{tab:oscillatory_params}.

\begin{table}[htbp]
\centering
\caption{Physical parameters of the resulting system. All these parameters were set by
trial and error and used for the simulations.}
\label{tab:oscillatory_params}
\begin{tabularx}{\textwidth}{CC}
\toprule
\textbf{Circuit Element} & \textbf{Physical Parameter} \\
\midrule
$C$                  & $1.0 \times 10^{-13}$ F \\
$R$                  & $2.0 \times 10^{3}$--$2.7 \times 10^{3}$ $\Omega$ \\
$R_{\mathrm{in}}$    & $1.0 \times 10^{4}$ $\Omega$ \\
$R_{c}$              & $1.2 \times 10^{6}$--$2.2 \times 10^{6}$ $\Omega$ \\
$V_{\mathrm{in}}$ (amplitude) & 1 V \\
\bottomrule
\end{tabularx}
\end{table}

In our setup, the frequency difference between neighboring oscillators should be around
20--40~MHz, so that a coherent, sinusoidal injected signal can synchronize them. The
precise value of the frequency difference depends on the strength of the input; the input
cannot be overly strong, however, because that would destroy the oscillatory behavior.

\subsubsection{Gradient-Free Optimization of Network Parameters}

To eliminate the somewhat heuristic procedure above, we used the Nevergrad package
\cite{ref:nevergrad} (version 1.0.8, written in Python) to let the algorithm determine the
parameters of the network, such as the optimal ring oscillator frequencies. For this
gradient-free optimization we used a genetic algorithm to find the frequencies and the
couplings.

\begin{figure}[htbp]
    \centering
    \includegraphics[width=0.92\textwidth]{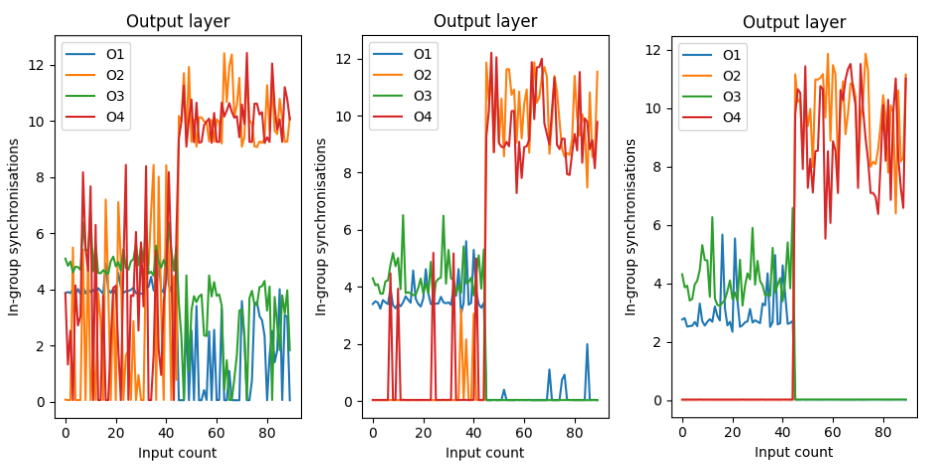}
    \caption{The synchronization value of the output pairs for the test dataset at
    different stages of learning. The initial state on the left is random; during learning
    the synchronization values start to settle towards the desired values; and finally,
    when learning is finished, the synchronization values indicate a 100\% separation.}
    \label{fig:nevergrad}
\end{figure}

Figure~\ref{fig:nevergrad} shows the same output-layer synchronization at different points
of the learning procedure. The leftmost panel is the initial state, the middle one an
intermediate step during learning, and the last one the state in which the network has
learned the $100\%$ separation.

The result is very similar to the trial-and-error, ``by hand'' solution, as the final
state of the learning indicates.

\subsection{Recognition of Raw Vowel Waveforms Without Preprocessing}
\label{sec:mixed}

In Figure~\ref{fig:vowel} and in Section~\ref{sec:voweleffect} we demonstrated that
applying the vowel waveform directly to an oscillator has a strong effect on its
frequency, even though this effect cannot be described as a clean phase and frequency
synchronization to an injected signal. The frequency correlations between neighboring
oscillators can nevertheless be picked up using appropriate circuitry.

\begin{figure}[htbp]
    \centering
    \includegraphics[width=\textwidth]{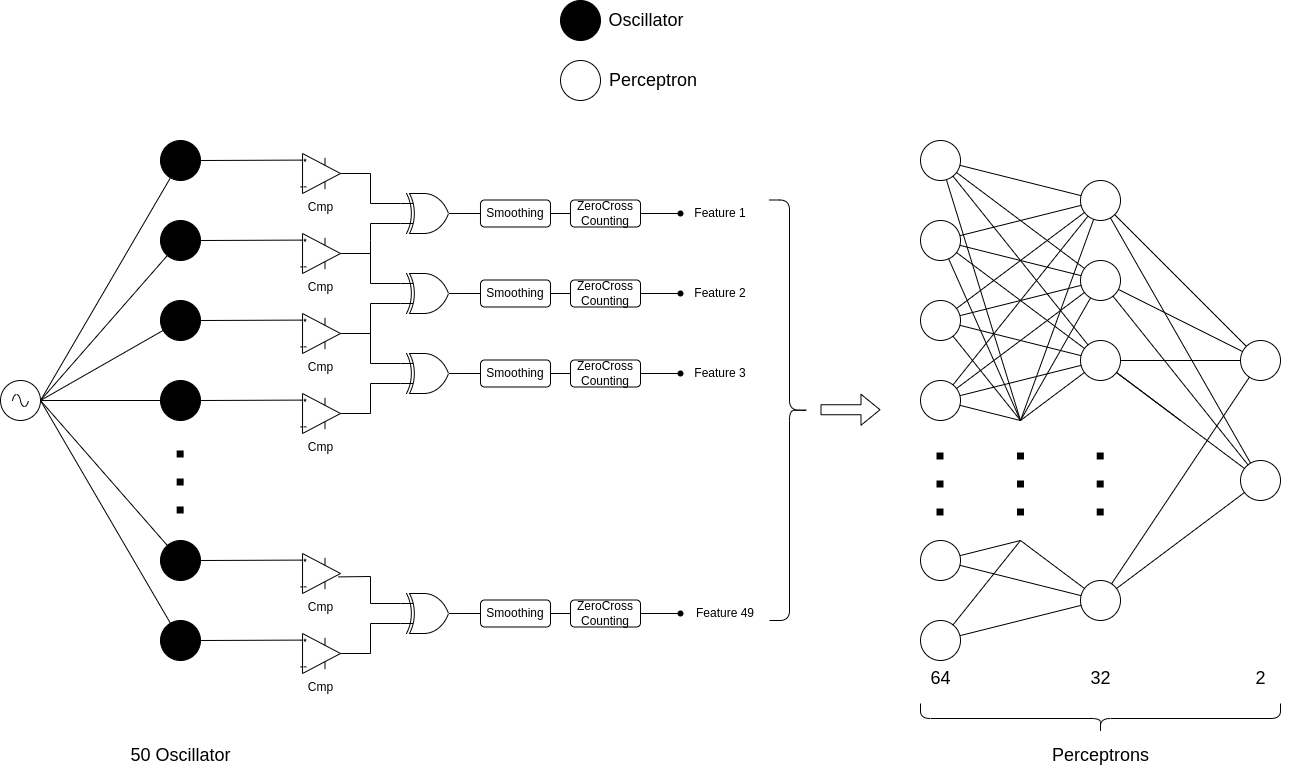}
    \caption{Mixed architecture for time-domain binary vowel recognition. The first part of
    the architecture extracts the frequency features of the time-domain vowels by measuring
    the frequency synchronization strength of neighboring oscillators. The second part uses
    this feature set as the input data for the actual binary recognition, performed by a
    simple multi-layer perceptron architecture.}
    \label{fig:mixed}
\end{figure}

In order to process raw waveforms, we use a single layer of oscillators with frequencies
spanning the spectral range of the vowels. Unlike the network of
Section~\ref{sec:multilayer}, this architecture uses a small conventional neural network
as a second layer to perform the classification; the oscillators act only as a
preprocessing layer. The architecture is shown schematically in Figure~\ref{fig:mixed}.

The first part of the architecture --- shown on the left side of Figure~\ref{fig:mixed}
--- is entirely oscillator-based. Its main function is to extract frequency components
from the time-domain vowel input by measuring the synchronization strength between
neighboring oscillators. Each oscillator's output voltage is first converted into a binary
signal using comparators. The outputs of adjacent oscillators are then passed through XOR
gates, which operate as follows:

\begin{itemize}[topsep=2pt,itemsep=1pt]
    \item if two oscillators are not synchronized in frequency, the duty cycle of their
          XOR output changes continuously;
    \item if two oscillators are synchronized in frequency, the duty cycle of their XOR
          output is constant over time.
\end{itemize}

The output of the XOR gates passes through a low-pass filter. This is straightforward to
realize in a circuit; in our simulations we used a moving-average filter to emulate its
effect.

\begin{figure}[htbp]
\centering
    \begin{minipage}[b]{0.48\textwidth}
        \centering
        \includegraphics[width=\linewidth]{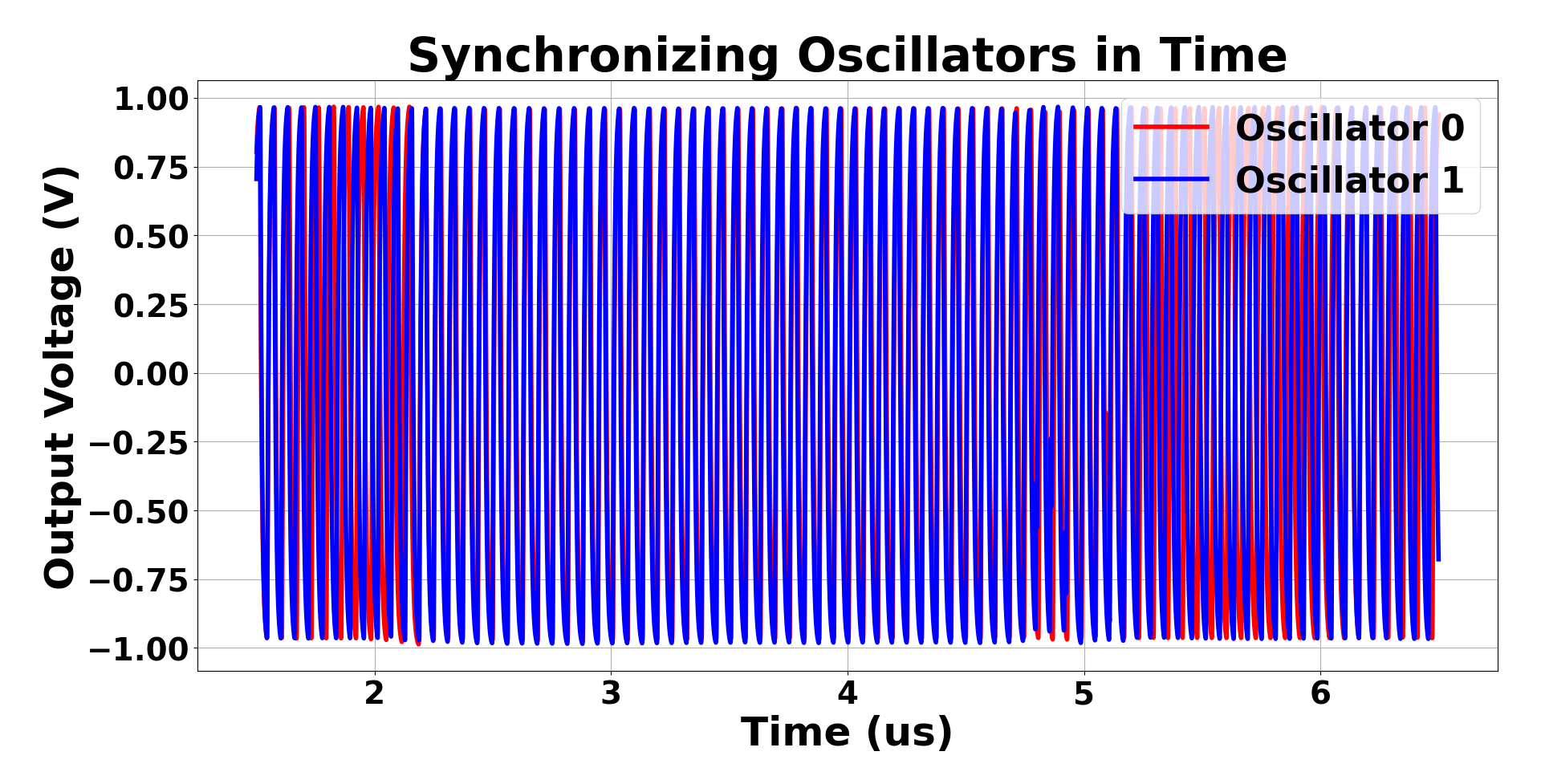}\\[2pt]
        (\textbf{a})
    \end{minipage}
    \hspace{0.2cm}
    \begin{minipage}[b]{0.48\textwidth}
        \centering
        \includegraphics[width=\linewidth]{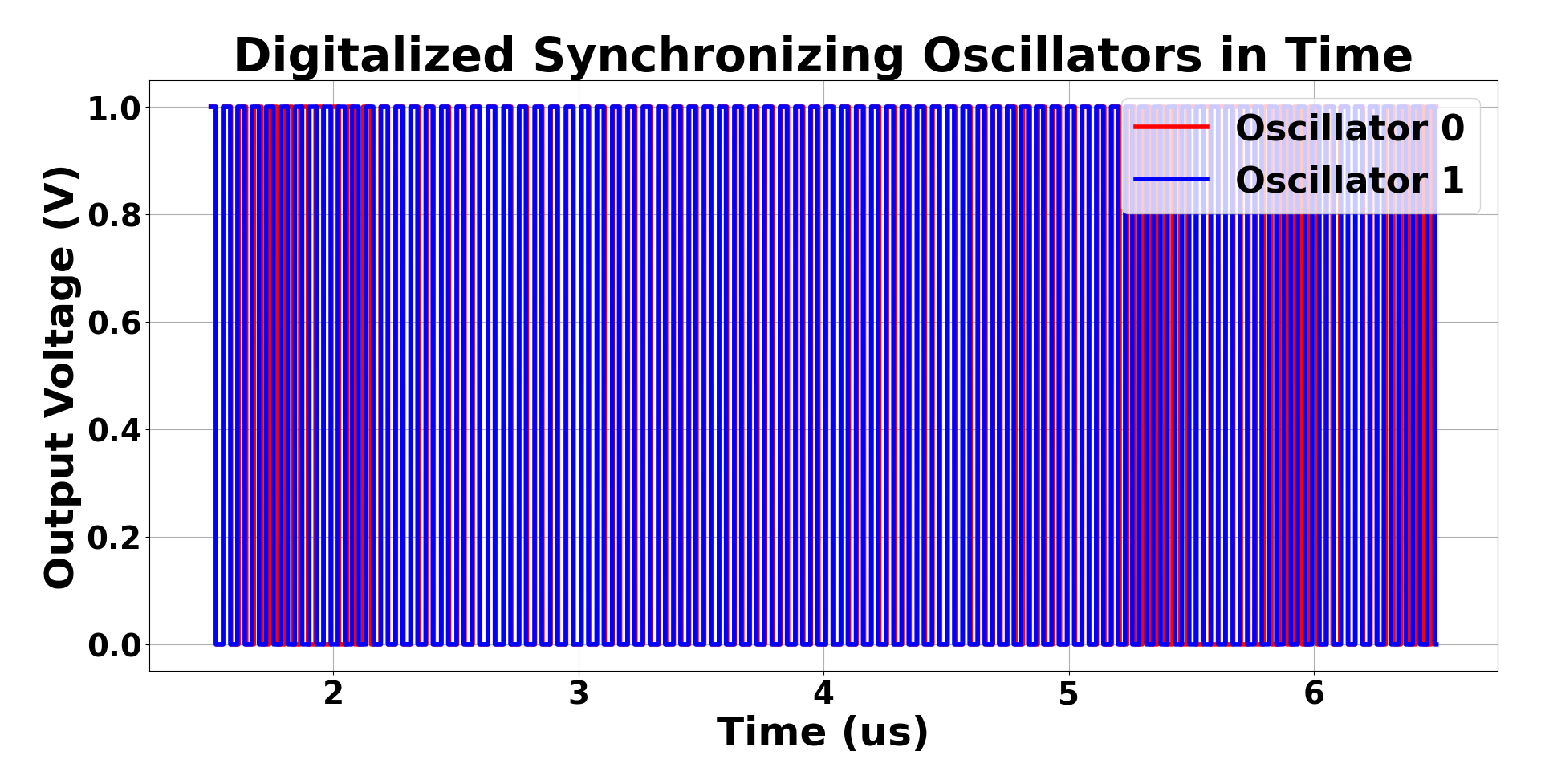}\\[2pt]
        (\textbf{b})
    \end{minipage}

    \vspace{0.4cm}

    \begin{minipage}[b]{0.48\textwidth}
        \centering
        \includegraphics[width=\linewidth]{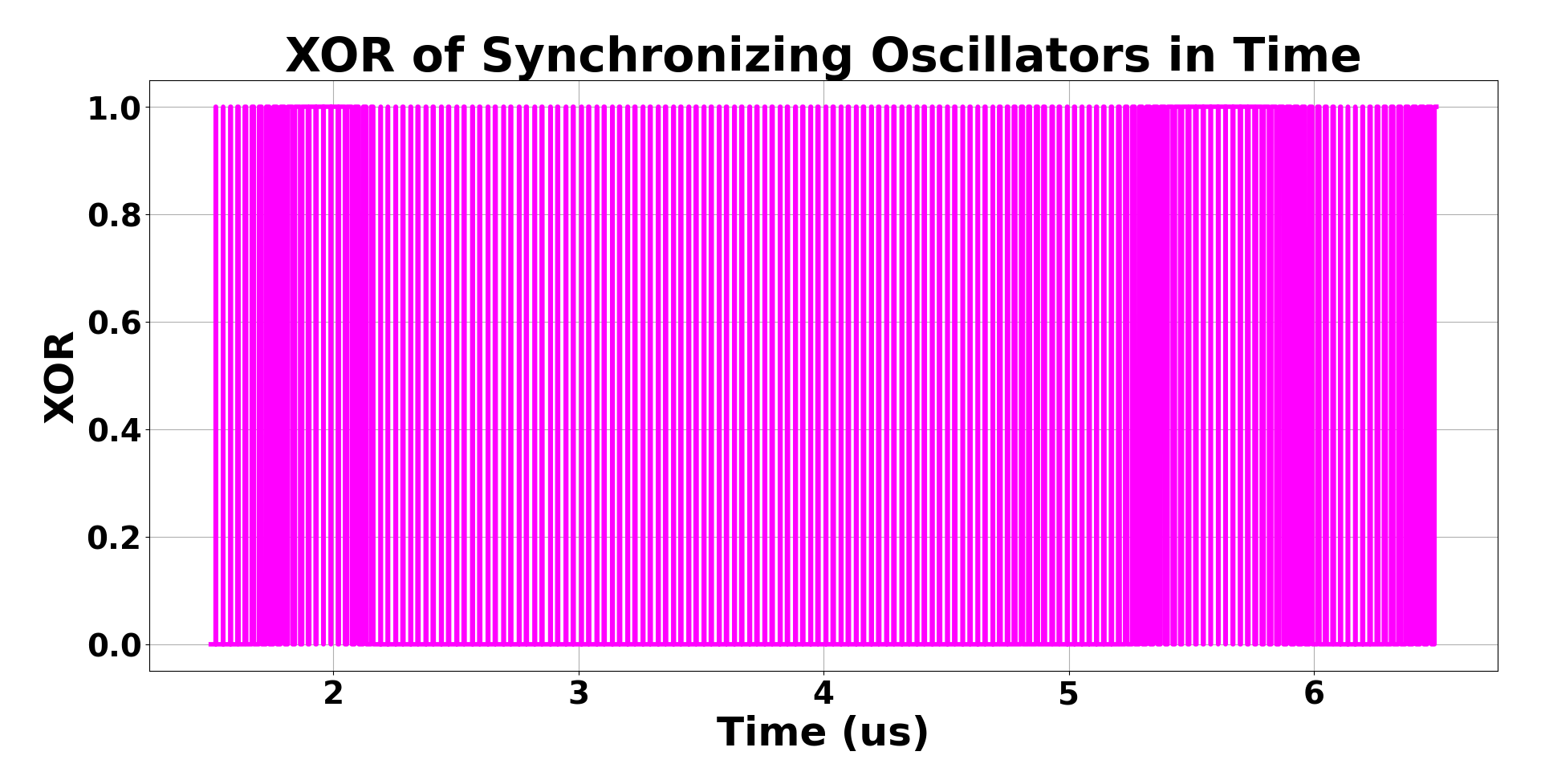}\\[2pt]
        (\textbf{c})
    \end{minipage}
    \hspace{0.2cm}
    \begin{minipage}[b]{0.48\textwidth}
        \centering
        \includegraphics[width=\linewidth]{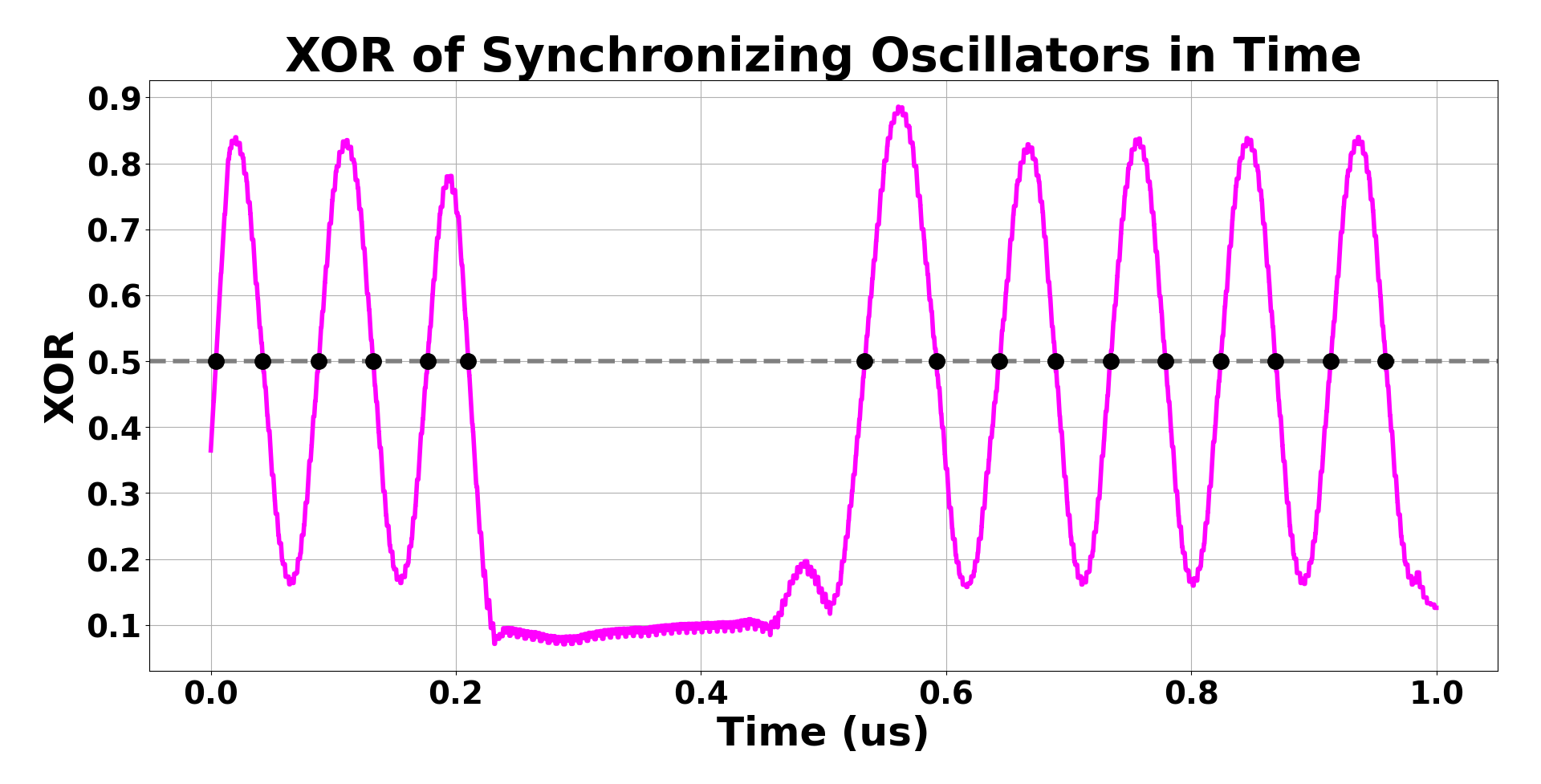}\\[2pt]
        (\textbf{d})
    \end{minipage}

    \caption{Transformations applied to the output signals of two neighboring,
    synchronizing oscillators by the small circuit that follows the oscillator layer.
    (\textbf{a}) Output voltages of the two oscillators around the synchronization time
    window: inside the synchronization region the frequencies of the two oscillators do
    not change relative to each other, whereas before and after it oscillator 1, being the
    higher-frequency one, drifts continuously with respect to oscillator 0.
    (\textbf{b}) The same signals after digitization: two square waves.
    (\textbf{c}) The XOR of the digitized signals, which shows behavior similar to
    (\textbf{a}): in the synchronization region the XOR has a constant duty cycle, while
    outside it the duty cycle changes continuously.
    (\textbf{d}) After applying a mean filter to the whole XOR signal, the synchronization
    region is clearly visible; counting the zero crossings of this signal gives a lower
    value than it would in the absence of synchronization, since outside the
    synchronization region the smoothed XOR signal is sinusoid-like. Note that we take the
    inverse of this value and map it to the $[0,1]$ range, which makes it more intuitive to
    call it a synchronization strength.}
    \label{fig:process}
\end{figure}

To create a single feature vector from the time-dependent filtered XOR signal, we count
its zero crossings. A constant duty cycle, indicating synchronized oscillators, gives few
zero crossings; a continuously changing duty cycle, indicating two non-synchronizing
oscillators, gives many. The successive transformations of the oscillator output voltages
as they pass through this apparatus are shown in Figure~\ref{fig:process}.

\begin{figure}[htbp]
    \centering
    \includegraphics[width=0.95\textwidth]{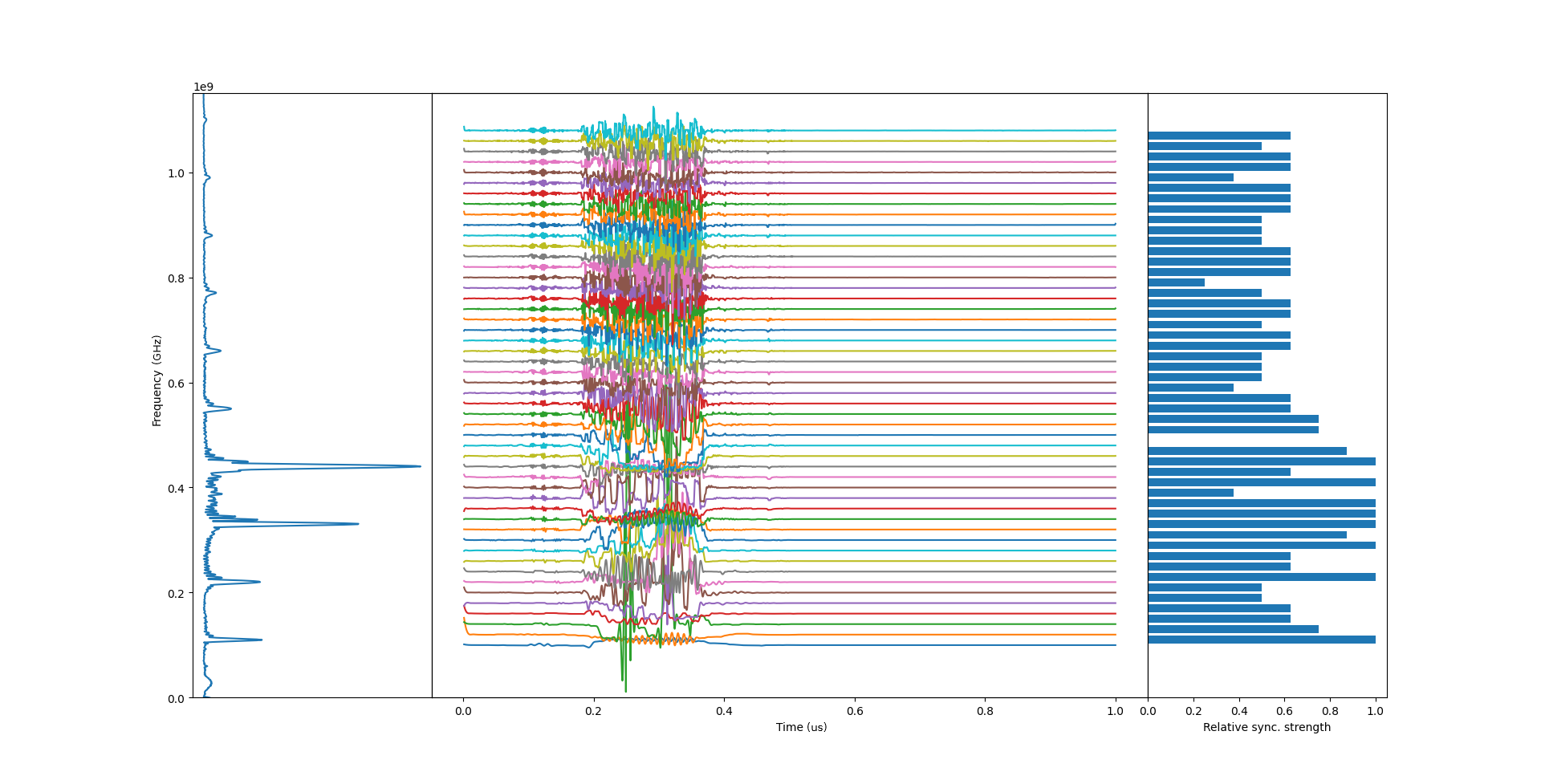}
    \caption{Example of feature extraction with the oscillatory layer. A feature-vector
    element is high if synchronization is present between two oscillators and low if there
    is no synchronization, as desired. The frequency spectrum of the vowel is shown on the
    left, rotated by 90 degrees for clarity. The colored curves are the instantaneous
    frequencies of the individual oscillators.}
    \label{fig:strength}
\end{figure}

An example of the oscillators' response to a time-dependent input is shown in
Figure~\ref{fig:strength}. It is clear from the figure that the oscillators with
free-running frequencies close to the frequency components of the vowel are synchronized,
while the rest are not. The right panel shows the relative frequency synchronization
strength between the oscillators: the zero-crossing counts have been normalized to the
$[0,1]$ interval and inverted, so that the frequency peaks of the vowel stand out more
clearly.

We first used the oscillatory architecture to perform the feature extraction on the two
vowel sets, and then used the resulting dataset as the input to the low-parameter-count
multi-layer perceptron mentioned above. It has $49$ inputs and $2$ outputs, with $64$
neurons in the first and $32$ neurons in the second hidden layer.

\subsubsection{Performance on the Dataset}

Because the dataset is small and this conventional neural network is simple, training and
evaluation of the architecture are fast.

\begin{figure}[htbp]
    \centering
    \includegraphics[width=0.55\textwidth]{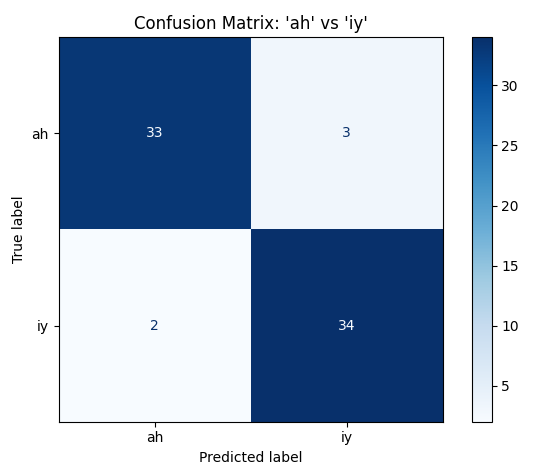}
    \caption{Confusion matrix for the ``ah'' and ``iy'' vowels with the mixed architecture.
    The architecture makes some errors, but its overall accuracy is $93\%$.}
    \label{fig:confusion}
\end{figure}

The confusion matrix for the evaluation of the test set is shown in
Figure~\ref{fig:confusion}.

In terms of accuracy this architecture performs slightly worse than the multi-layer
oscillatory network, but there the input signals were heavily preprocessed, whereas here
the preprocessing is done by the oscillatory layer itself. Note also that the multi-layer
perceptron is a very simple network and is therefore prone to overfitting, especially on
the small dataset available to us.

As for the physical parameters, all the oscillators had the same input resistor value of
$2.5$~k$\Omega$. The remark at the end of Section~\ref{sec:params} applies here as well,
with the added restriction that the difference between the frequencies of neighboring
oscillators should be at most $20$--$25$~MHz, so that two to four oscillators can
synchronize in frequency to the incoherent input signal for the duration of the speech.

%=====================================================================
\section{Discussion}

We presented ONN architectures for processing time-domain vowel signals.

Our first architecture consists solely of oscillators, but its incoming signals require
some preprocessing: it receives formants (coherent oscillatory signals) extracted from the
vowels. If the oscillator frequencies are properly chosen, the network can achieve 100\%
classification accuracy on simple vowels, correctly identifying their frequency
constituents.

Classification therefore depends largely on the separability of the formant frequencies,
especially the two most dominant ones. Although we did manage to reach 100\% accuracy for
some classes, others performed worse, around 80--90\%, depending on the linear separability
of the formant frequencies of the given vowels.

A key feature of this network is that it has two oscillator layers; frequency-domain
information is passed from the first layer to the second, akin to a feed-forward network.
We envision that this architecture can be developed to perform more complex processing
functions by making the connections of the second layer more complex.

The second architecture is a single-layer ONN that receives entirely unprocessed signals,
i.e., raw waveforms. For an injected broadband signal, a degree of entrainment occurs
between the oscillators. This synchronization cannot be passed on to another oscillator
layer, but it can be used to extract a feature vector by means of simple circuitry, which
is then processed further by a simple, more conventional neural network.

For this architecture the accuracy was lower than for the first architecture on the test
case shown in this paper, but we emphasize that the first approach uses a preprocessed,
coherent signal, while this solution uses the raw waveform. Compared with the 100\% of the
first architecture on the ``ah'' and ``iy'' vowels, we achieved 93\% on the test set,
which is high for such simple circuitry.

We also tried multi-class classification with the second architecture, using $12$ output
neurons instead of $2$, and obtained good results on the training and validation sets,
reaching $97\%$. This means that the oscillatory layer managed to extract meaningful
information from the raw signals, essentially transforming a time-dependent signal into
static data. It should be mentioned, however, that the network was overfitting and in some
cases failed to generalize to the test set, mainly because of the small number of samples
per class in the test set.

We believe that the overfitting was likely due to the simplicity of the traditional second
layer, which is still an extremely simple and lightweight neural network. Work is in
progress to optimize the performance of the non-oscillatory layers of the architecture.

The motivation for this work is to achieve ultra-low-power signal processing.
Ring-oscillator-based ONNs have been shown to outperform the power efficiency of their
digital counterparts by two orders of magnitude for simple tasks \cite{ref:frontiers}. The
ONN presented here runs at a frequency of several hundred MHz, can perform several tens of
thousands of classifications per second, and consists of a few tens of ring oscillators,
each of which may consume a mere few microwatts of power \cite{ref:ro}; the energy cost per
inference therefore falls into the nJ regime. This is in line with state-of-the-art
hardware classifiers for static data \cite{ref:bench} and significantly better than the
state of the art for preprocessing-hungry temporal processing.

ONNs are most commonly used in image processing tasks, typically as associative memories.
In contrast, processing dynamic signals remains a relatively underexplored area, despite
ONNs being inherently dynamic systems themselves and therefore potentially well suited to
such tasks. We hope that our work encourages further research into the potential of ONNs
for dynamic signal processing.

%=====================================================================
\section*{Author Contributions}

Conceptualization and funding acquisition: G.C. and W.P. Software, validation, data
analysis and concept development: T.R.-H. Writing --- original draft preparation: T.R.-H.
and G.C. All authors have read and agreed to the published version of the manuscript.

\section*{Funding}

The authors declare that financial support was received for the research, authorship
and/or publication of this article. This study was partially supported by a grant from
Intel Corporation, titled HIMON: Hierarchically Interconnected Oscillator Networks. This
project has received additional funding from the European Union's Horizon EU research and
innovation programme under grant agreement No.\ 101092096 (PHASTRAC). The funder (Intel
Corporation) had the following involvement with the study: helping to discover relevant
literature and motivate the study. The funder was not involved in the study design, the
collection, analysis or interpretation of data, the writing of this article, or the
decision to submit it for publication.

\section*{Data Availability}

The vowel dataset used by the authors can be found at
\url{https://github.com/fancompute/wavetorch} and was downloaded in January 2024.

\section*{Acknowledgments}

The authors are grateful for useful discussions with Amir Khosrowshahi, Narayan Srinivasa
and Dmitri Nikonov --- all at that time at Intel Corporation.

\section*{Conflicts of Interest}

The authors declare no conflicts of interest.

%=====================================================================

\end{document}